\documentclass[conference]{IEEEtran}
\IEEEoverridecommandlockouts

\usepackage{amsmath,amssymb,amsfonts}
\usepackage[ruled]{algorithm2e}
\usepackage{graphicx}
\usepackage{textcomp}
\usepackage{xcolor}
\def\BibTeX{{\rm B\kern-.05em{\sc i\kern-.025em b}\kern-.08em
    T\kern-.1667em\lower.7ex\hbox{E}\kern-.125emX}}

\usepackage[OT1]{fontenc}

\usepackage[style=ieee,maxbibnames=9]{biblatex}
\usepackage{mathtools}
\usepackage{bm}
\usepackage{subcaption}

\usepackage{booktabs}

\usepackage{mwe}

\usepackage[draft, commentmarkup=todo,
  todonotes={textsize=tiny, textwidth=0.83in}]{changes}

\definechangesauthor[name={Ilya}, color=red]{is}

\DeclareMathOperator*{\argmax}{argmax}

\usepackage{mathtools}

\begin{document}

\title{Scaling Up the Quantum Divide and Conquer Algorithm for Combinatorial Optimization\\

\thanks{This project was funded in part by (1) DARPA under the ONISQ program and (2) QNEXT the U.S. Department of Energy, Office of Science, National Quantum Information Science Research Center.} 
}

\makeatletter
\newcommand{\linebreakand}{%
  \end{@IEEEauthorhalign}
  \hfill\mbox{}\par
  \mbox{}\hfill\begin{@IEEEauthorhalign}
}
\makeatother

\author{\IEEEauthorblockN{Cameron Ibrahim}
    \IEEEauthorblockA{\textit{Computer \& Information Sciences} \\
        \textit{University of Delaware}\\
        Newark, DE \\
        cibrahim@udel.edu}
    \and
    \IEEEauthorblockN{Teague Tomesh}
    \IEEEauthorblockA{
        \textit{Infleqtion}\\
        Chicago, IL \\
        teague.tomesh@infleqtion.com}
    \linebreakand
    \IEEEauthorblockN{Zain Saleem}
    \IEEEauthorblockA{\textit{Mathematics and Computer Science Division} \\
        \textit{Argonne National Laboratory}\\
        Lemont, IL\\
        zsaleem@anl.gov}
    \and
    \IEEEauthorblockN{ Ilya Safro}
    \IEEEauthorblockA{\textit{Computer \& Information Sciences} \\
        \textit{University of Delaware}\\
        Newark, DE \\
        isafro@udel.edu}
}

\maketitle

\begin{abstract}
    Quantum optimization as a field has largely been restricted by the constraints of current quantum computing hardware, as limitations on size, performance, and fidelity mean most non-trivial problem instances won't fit on quantum devices. Even proposed solutions such as distributed quantum computing systems may struggle to achieve scale due to the high cost of inter-device communication. To address these concerns, we propose Deferred Constraint Quantum Divide and Conquer Algorithm (DC-QDCA), a method for constructing quantum circuits which greatly reduces inter-device communication costs for some quantum graph optimization algorithms. This is achieved by identifying a set of vertices whose removal partitions the input graph, known as a separator; by manipulating the placement of constraints associated with the vertices in the separator, we can greatly simplify the topology of the optimization circuit, reducing the number of required inter-device operations. Furthermore, we introduce an iterative algorithm which builds on these techniques to find solutions for problems with potentially thousands of variables. Our experimental results using quantum simulators have shown that we can construct tractable circuits nearly three times the size of previous QDCA methods while retaining a similar or greater level of quality.

\end{abstract}


\section{Introduction}

Quantum advantage describes the potential speedup achievable using quantum computers on tasks traditionally performed on classical hardware, such as cryptography \cite{shor1994algorithms}, machine learning \cite{beer2020training}, and optimization \cite{herman2023quantum}. However, many relevant problems in these areas are fairly tractable for small inputs using classical methods. In order to demonstrate an advantage, we must push the boundary on the size of inputs that we can solve using quantum computers, which are currently constrained in both size and error rate \cite{preskill2018quantum}.

Distributed quantum computing systems may provide the solution to the issue of scalability \cite{cuomo2020towards}.  Since the error rate of a quantum device tends to inflate as its size increases, advocates of distributed quantum computing systems propose to utilize a variety of smaller quantum devices connected by means of quantum teleportation protocols. In the short term, implementation of these systems would be constrained by the cost of inter-device operations, which can add significant overhead and decrease fidelity depending on what method is used \cite{hu2023progress, ayral2021quantum, ayral2020quantum}. In order to successfully study the scalability of these distributed quantum systems, we must then find ways to work around this overhead.

Classical distributed systems similarly incur a significant overhead on inter-device operations, which may lead to a bottleneck in system performance \cite{woodruff2017distributed}. Systems engineers generally try to limit this bottleneck by mapping processes to individual devices to minimize the necessity of inter-device operations (e.g., this can be done by modeling communication complexity using graph partitioning \cite{Buluc2016}). However, this is only necessary insofar as the inter-device operations present a bottleneck; once below a certain threshold, it becomes useful to instead try to optimize for device utilization, ensuring that the task at hand is making efficient use of the resources in the distributed system.

These classical techniques inspired methods such as the Quantum Divide and Conquer Algorithm (QDCA), which attempts to decompose a single combinatorial optimization problem into subproblems which can be solved on individual quantum devices \cite{dqca}. This decomposition models the input as a graph, and attempts to minimize the amount of information which must be passed between devices by applying graph vertex partition algorithms to the underlying graph topology. In the QDCA approach, the problem is solved by applying a series of quantum gates known as mixers to each variable of the problem. If the number of inter-device operations is still prohibitive, the problem is sparsified by deactivating a subset of these mixers in order to further reduce the number of inter-device operations below a tractable threshold. 

The benefits of QDCA have previously been shown while working with the Maximum Independent Set problem. This is a classical NP-Hard graph optimization problem with a wide variety of applications. 
This is a well studied problem in quantum optimization \cite{tomesh2022quantum}, and will be the primary area of focus for this paper.

\noindent {\bf Our contribution} In order to address the limitations of QDCA with regards to scalability and device utilization, we introduce the Deferred Constraint Quantum Divide and Conquer Algorithm  (DC-QDCA). Our approach utilizes edge partitioning in order to compute a k-way separator for our input graph. By manipulating the position of constraints associated with this separator, we can construct a circuit whose simplified topology allows it to decomposed into \(k+1\) subcircuits. The first \(k\) of these subcircuits are completely disconnected from one another and can be run entirely in parallel with no required inter-device operations.  The final circuit corresponds to our separator; for each vertex adjacent to the separator, we must communicate its result to the separator circuit. We can adjust the required number of inter-device operations solely by deactivating mixers in the separator circuit, leaving each of the other \(k\) subcircuits completely active. 

Utilizing this approach, we are able to significantly increase the number of vertices participating in the optimization of a single circuit for a given budget of inter-device operations, increasing the quality of the result. Moreover, for very large graphs where it is infeasible to produce a connected circuit with this approach, we introduce a method which iterates over individual vertices in the separator and attempts to solve the subgraph of partitions connected to that node. Using this approach, we can acquire quality results for graphs with several thousand vertices. Our approach directly benefits the distributed quantum computing algorithms by scaling them up.

\section{Background  \& Related Work}

\subsection{The Maximum Independent Set Problem}

An undirected graph \(G = (V, E)\) with no loops and multi-edges is a set \(V\) of vertices together with a set \(E \subseteq \binom{V}{2}\) of edges. 
An edge \(e\) and a vertex \(v\) are incident with one another if \(v \in e.\) Two vertices \(u, v\) are adjacent if there exists an edge \(e\) such that \(u,v\in e.\) The neighborhood \(N(v)\) of a vertex \(v\) is the set of vertices adjacent to \(v.\)

An independent set is a subset \(S \subseteq V\) where no two vertices \(u,v \in S\) are adjacent. An independent set \(S\) is maximal if for all \(v \in V\), \(S \cup \{v\}\) is not an independent set. The maximum independent set for a graph \(G\) is the maximal independent set with largest cardinality. 

The Maximum Independent Set Problem admits a formulation as the following binary linear program:
\begin{align*}
    \max\quad      & \sum_{v\in V} x_v,                     \\
    \text{such that}\quad & x_u + x_v = 1          & \forall uv \in E, \\
                   & x_v \in \{0, 1\}       & \forall v \in V.
\end{align*}

\subsection{Vertex \& Edge Partitioning}

Graph partitioning is generally synonymous with graph vertex partitioning, where the vertex set \(V\) of the graph is split into $k$ disjoint sets \(V_1 \sqcup \cdots \sqcup V_k.\) For \(k=2\), this is commonly known as Graph Bisection. A vertex partitioning solver will generally try to minimize the number of edges crossing between individual partitions. In balanced vertex partitioning, a constraint is added guaranteeing that no one partition takes too great a portion of the graph. The full formulation of balanced vertex partitioning is given by
\begin{align*}
    \min_{V = V_1 \sqcup \cdots \sqcup V_k}\qquad & \lvert \{ uv \in E \mid \exists i\neq j :  v \in V_i \wedge u \in V_j \} \rvert                                                                  \\
    \text{st}\qquad & \forall i,  \vert V_i\rvert \leq (1 + \epsilon) \left\lceil \frac{\lvert V \rvert}{k}\right\rceil.
\end{align*}

By comparison, graph edge partitioning attempts to split the edge set \(E\) into disjoint sets \(E_1 \sqcup \cdots \sqcup E_k.\) Edge partitioning has seen success in process assignment in classical distributed systems for systems which a few processes with a large number of required interactions with other processes \cite{gonzalez2012powergraph}. Dual to the notion of an edge partition is the idea of a vertex separator, the set of vertices \(S\) which are incident to at least two edges assigned to different partitions. 

\begin{figure}[t]
    \centering
    \includegraphics[width=0.20\textwidth]{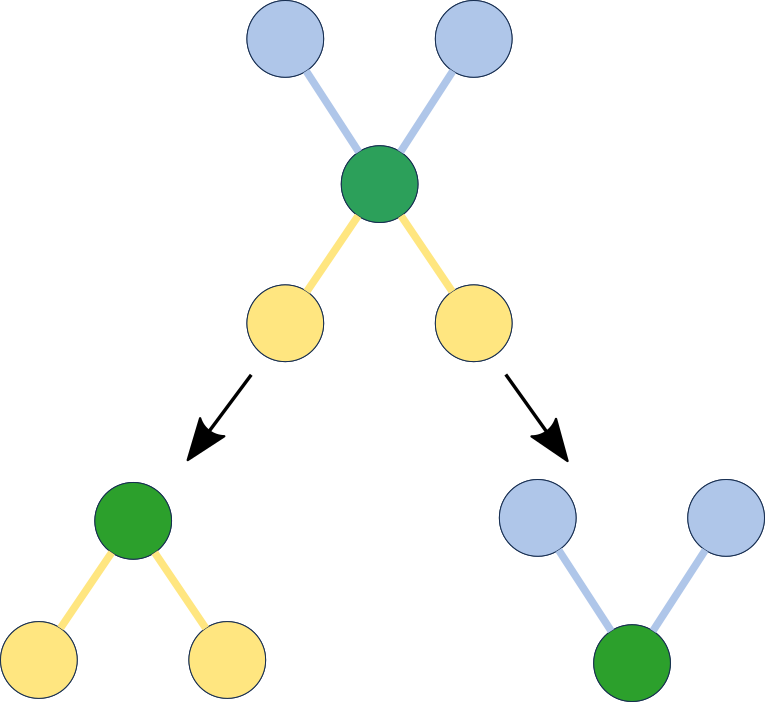}
    \caption{An edge partition assigns a label to each edge of a graph \(G\). The set of vertices with edges belonging to multiple partitions (green) form a separator for the rest of the graph.}
    \label{fig:enter-label}
\end{figure}

A vertex separator induces a vertex partition \(V = V_1 \sqcup \cdots \sqcup V_k \sqcup S,\) where every edge in the graph either has one end in the separator or both ends in a single partition.  An edge partitioning solver will generally try to find a solution which minimizes the sum of the number of edge partitions each vertex in the set is incident to. An analogous balance constraint can also be defined, giving the final formulation
\begin{align*}
    \min_{E = E_1 \sqcup \cdots \sqcup E_k}\qquad & \sum_{v \in V} \lvert \{ i \mid \exists u \in N(v) : uv \in E_i \} \rvert - 1                                                                  \\
    \text{such that}\qquad & \forall i,  \vert E_i\rvert \leq (1 + \epsilon) \left\lceil \frac{\lvert E \rvert}{k}\right\rceil.
\end{align*}

\subsection{Circuit Cutting and Knitting}

A distributed quantum computing (DQC) system consists of a collection of smaller quantum computing devices which are networked together via inter-device operations \cite{caleffi2022distributed}. In this way, they can have a large number of effective qubits, without the negative effects of increasing the number of qubits within a single device, such as increased error rates \cite{tomesh2022supermarq}.

In order to map a circuit designed for a single device onto a DQC,  we utilize a technique known as circuit cutting. This method decomposes a single circuit into a number of subcircuits which will be mapped to individual devices in the DQC; the results of one subcircuit will be correlated with the initial state of another, communicated using inter-device operations \cite{peng2020simulating, piveteau2023circuit, mitarai2021overhead}. 
These input-output pairings can be represented as cuts on the wires or gates of the quantum circuit. 
Dual to the notion of circuit cutting is circuit knitting, where individual subcircuits are recombined to acquire an overall solution \cite{piveteau2023circuit}.

There is a considerable cost associated with mapping single device circuits to a DQC. In particular, without quantum methods of communicating between devices such as teleportation, the cost of classical methods for reconstructing the result of a single circuit evaluated on a DQC is exponential in the number of cuts required to decompose the circuit \cite{piveteau2023circuit}. This places considerable constraints on what circuits can currently be evaluated using these methods.

\subsection{QAOA and Quantum Divide and Conquer}

Maximum Independent Set can similarly be formulated in terms of quantum operations. This problem has previously been studied in the context of hybrid quantum-classical optimization, where classical optimization methods such as gradient descent are used to optimize the parameters of a quantum circuit which outputs a valid solution to the given problem \cite{zhou2020quantum, saleem2023approaches}. These types of parameterized circuits are generally called \textit{ansatz}.

In this setting, the Maximum Independent Set objective is known as the Hamming weight of the solution bitstring \(x \in \{0,1\}^{\lvert V \rvert}\).
The Hamming Weight function can be represented in terms of quantum operations as
\begin{align*}
    C = \sum_{v \in V} \frac{1 - Z_v}{2},
\end{align*}
where \(Z_v\) is the Pauli-Z operator applied to the qubit corresponding to vertex \(v\) \cite{dqca}.
For an ansatz \(\psi\) with parameters \(\theta\), the loss function which is minimized in order to find an optimal independent set is given by the expectation
\begin{align*}
    \mathcal{L}(\theta) = -\mathbb{E}\Big[\langle \psi(\theta) \vert C \vert \psi(\theta)\rangle\Big].
\end{align*}

The Quantum Approximate Optimization Algorithm (QAOA) is a class of algorithms for combinatorial optimization which follow this ansatz construction approach \cite{zhou2020quantum}. The ansatz in QAOA is generally constructed using an initial state \(\lvert s\rangle\) and \(p\) alternating layers of parametrized phase unitaries \(U_C(\theta)\) and mixing unitaries \(U_M(\theta)\)
\begin{align*}
    \lvert \psi(\theta) \rangle = U^p_C(\theta)U^p_M(\theta)\cdots U^1_C(\theta)U^1_M(\theta)\lvert s \rangle,
\end{align*}
where 
\begin{align*}
    U_C^k(\theta) &= \exp\{-i\theta_{C,k} C\}, & U_M^k(\theta) &= \exp\{-i\theta_{M, k} M\}
\end{align*}
for some problem dependent operator \(M\). Accelerating QAOA and making it more practical is of great importance \cite{ushijima2021multilevel,galda2similarity}.

In this paper, we  build upon a specific variant of QAOA known as QDCA. In QDCA, the mixing unitaries \(U_M\) are defined such that they enforce the constraints of the relevant combinatorial optimization problem \cite{saleem2023approaches}. The mixing unitary \(U_M(\theta)\) is defined such that \(U_M^k(\theta) = \prod_{v \in V} P_v(\theta_{M,k,v})\), where each \(P_v(\alpha) = \exp\{-i\alpha M_v\}\) is a partial mixer on the wire corresponding to the vertex \(v.\) For Maximum Independent set, \(M_v\) is defined as 
\begin{align*}
    M_v &\coloneqq X_v \prod_{\mathclap{u \in N(v)}} \frac{1 + Z_u}{2},
\end{align*}
where \(X_v\) is the Pauli-X operator on the qubit corresponding to the vertex \(v\). In words, \(M_v\) will flip \(x_v\) if and only if none of its neighbors are in the \(\lvert 1 \rangle\) state.

Furthermore, QDCA incorporates both vertex partitioning and circuit knitting to scale to larger graphs. The QDCA algorithm utilizes graph bisection to acquire a partition of the vertex set \(V = V_1 \sqcup V_2.\) Individual ansatz are constructed for each of the induced subgraphs corresponding to these partitions. Vertices which are incident to an edge which crosses
between these two vertex sets are referred to as cut nodes, and are given by the set \(Q \subseteq V\). A certain number of ``hot'' nodes are then chosen from \(Q;\) these hot nodes are used to knit the constructed ansatz together to allow gradient information flow between the two circuits. The remaining ``cold'' nodes have their mixers deactivated, removing them from consideration for the solution for this circuit so as to remove the possibility of an invalid solution. In this way, decreasing the number of ``hot'' nodes decreases the number of required inter-device operations at the cost of decreased solution quality.

\section{Deferred Constraint Divide and Conquer}

In this section, we attempt to alleviate the issue of mixer deactivations in order to improve the quality-scalability tradeoff in QDCA. To do so, we introduce the Deferred Constraint Quantum Divide and Conquer Algorithm  (DC-QDCA). We do so in order to simplify the structure of the circuit, which in turn allows for a greater number of active mixers.

The power of the deferred constraint approach lies in the simplicity with which we can compute its overhead. The number of wire cuts required to split a deferred constraint circuit is exactly equal to the size of the neighborhood of the separator, \(\bigcup_{v\in S} N(v) \setminus S\). By trying to minimize this quantity, we minimize the number of cuts required to separate the circuit. Once a separator is chosen, we have the option to sparsify it, removing vertices until the number of required cuts is within our budget. This process is equivalent to deactivating the mixers corresponding to those vertices in the separator.

\subsection{Construction of the QC-QDCA Circuit}

\begin{figure}[t]
    \centering
    \begin{subfigure}{0.45\textwidth}
        \centering
        \includegraphics[width=0.6\textwidth]{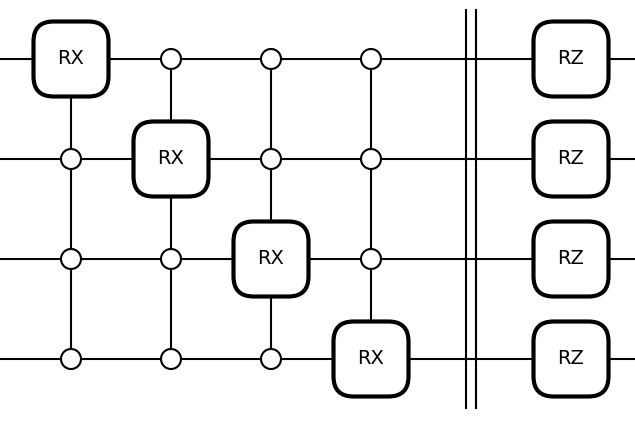}
        \caption{}
        \label{extra-mixers}
    \end{subfigure}

    \begin{subfigure}{0.45\textwidth}
        \centering
        \includegraphics[width=0.6\textwidth]{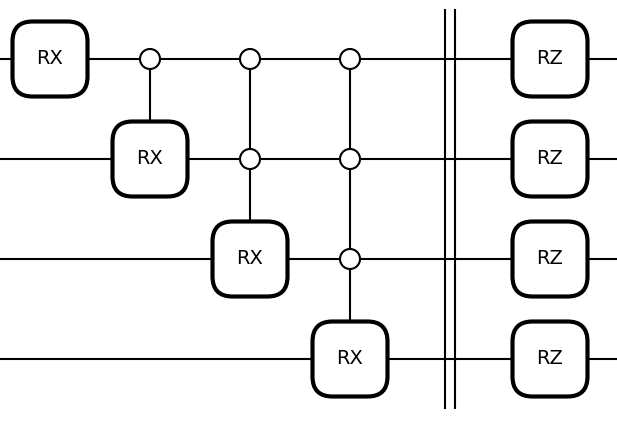}
        \caption{}
        \label{simple-mixers}
    \end{subfigure}
    \caption{Example of circuit simplification. Circuit (a) is simplified into (b) by removing unnecessary connections.}
    \label{fig:simplification}
\end{figure}

A key insight enabling DC-QDCA is that if a vertex \(v\) has a neighbor \(u\), where \(u\) was not in the independent set in the initial state and \(u\) has not yet had a partial mixer applied to it, then a partial mixer on \(v\) does not have to check the current state of \(u\). Given an ordering \(v_1, \ldots, v_{\lvert V\rvert} \in V\) of the vertices of the input graph which matches the ordering of partial mixers in the circuit, this can be expressed in terms of gate operations as 
\begin{align*}
    M_v &\coloneqq X_v \prod_{\mathclap{v_j \in \mathcal{N}} } \frac{1 + Z_{v_j}}{2}, & \mathcal{N} \coloneqq \{ v_j \in N(v_i) \mid j < i \}.
\end{align*}
This is reflected in Fig. \ref{fig:simplification}, where for an initial state corresponding to the empty set, the partial mixer for each vertex only needs to check the wires corresponding to neighbors coming before it in the order.

Without circuit cutting, this benefit may seem negligible. However, this reduction in the number of connections between wires can significantly reduce the number of required inter-device operations. Consider a vertex separator \(V = V_1 \sqcup \cdots \sqcup V_k \sqcup S\) of the input graph. Each edge of the graph is either contained completely in one partition, has at least one end in \(S\). If the mixers corresponding to the vertices in \(S\) are placed at the end of the circuit, then no gate earlier in the circuit will need to connect two wires from different partitions. As a result, if we identify partition subcircuits \(\psi_i\) corresponding to each partition \(V_i\), as well as a separator subcircuit \(\psi_S\) corresponding to \(S\), no inter-device operations will need to be used between any two \(\psi_i, \psi_j\). This is construction is demonstrated in Fig. \ref{deferred-mixers}, where deferring the separator produces a circuit with three subcircuits. Notably, this construction also means that each of the partition subcircuits can be executed completely in parallel, even on a classical simulator.

\begin{figure}[t]
    \centering
    \begin{subfigure}{0.15\textwidth}
        \centering
        \includegraphics[width=\textwidth]{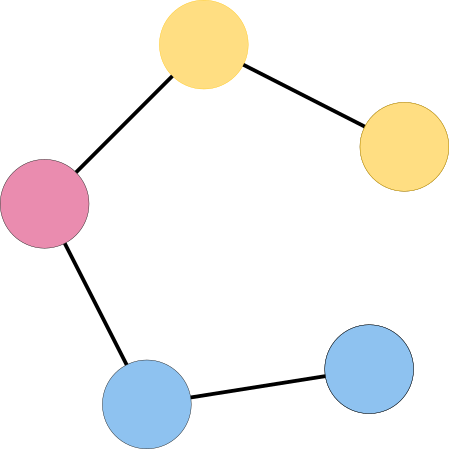}
        \caption{5 node path graph}
    \end{subfigure}\\
    \begin{subfigure}{0.35\textwidth}
        \centering
        \includegraphics[width=\textwidth]{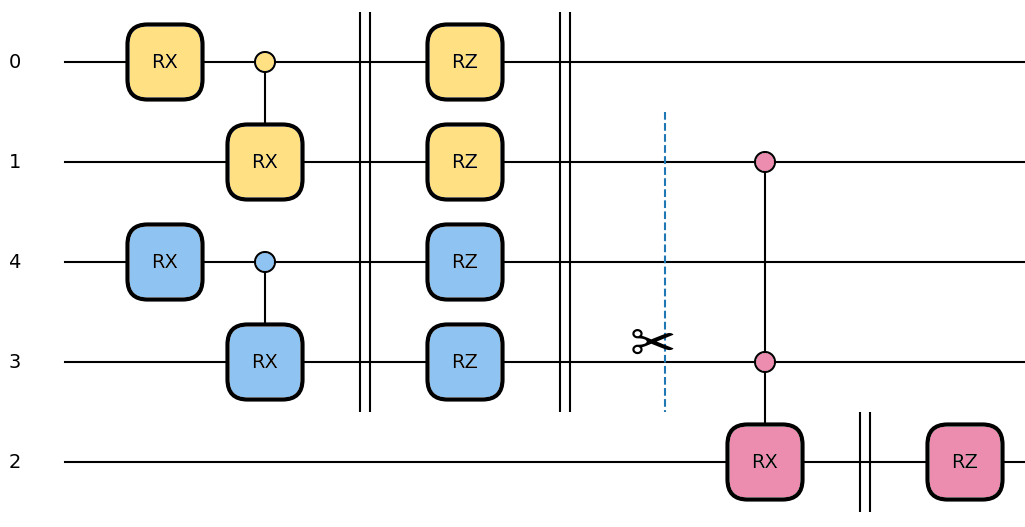}
        \caption{A deferred constraint circuit.}
    \end{subfigure}
    \caption{Example using a vertex separator (Red) with two partitions (Blue, Yellow). The DC-DQCA circuit is separated into 3 subcircuits using cuts on wires 1 and 3, corresponding to nodes adjacent to the separator.}
    \label{deferred-mixers}
\end{figure}

The number of cuts required to split the circuit in this way is then going to be \(\bigcup_{v\in S} N(v) \setminus S\), which is the number of vertices not in the separator whose value must be communicated to the separator circuit. Each individual subcircuit \(\psi\) will be constructed as in QDCA, but using the simplified mixers defined in this section. The overall DC-QDCA circuit has the form
\begin{align*}
    \psi(\theta) &\coloneqq \psi_S(\theta)\prod_{i=1}^k \psi_i(\theta).
\end{align*}

In order to further simplify the circuit, we order the partial mixers of the graph in order of neighborhood size, \(\lvert N(v)\rvert.\) This is a criterion used in some greedy algorithms for Maximum Independent set, and leads to an an order in which there few vertices with neighbors earlier in the order \cite{kako2009approximation}. When combined with the simplified mixers defined in this section, this can in practice reduce the number of wire connections within each subcircuit.

\subsection{Separator Sparsification}

While the DC-QDCA algorithm produces a circuit requiring only a small number of cuts, this can still quickly become intractable due to the exponential cost of overhead associated with inter-device-communication. As such, we require some way to simplify the problem so that it may still be tractably evaluated. In this section, we present two approaches to achieving tractability, one through sparsification and one through iteration.

As the number of cuts required to split the QC-QDCA circuit is given \(\bigcup_{v\in S} N(v) \setminus S\), one approach to decreasing this cost is to remove vertices, deactivating their associated mixers, in order to decrease the size of the neighborhood adjacent to the separator. In this paper, we choose a subset of the separator by fixing an ordering of the vertices in the separator and iteratively attempting to add them to a candidate set. By maintaining the current neighborhood of the candidate set, rather than evaluating the neighborhoods of each vertex  independently, we are able to capture vertices with overlapping neighborhoods that are not going to increase the number of required cuts. We once again utilize an ordering based on neighborhood size in order to prioritize vertices with few connections to the rest of the graph. This approach is appropriate when only a few vertices need to be removed to produce a tractable circuit, one which can solve the input problem using a single circuit. This approach is summarized as Alg. \ref{alg:circuit}.

\begin{algorithm}
    \caption{Deferred Constraint Quantum Divide and Conquer Algorithm (Single Circuit)}\label{alg:circuit}
    \DontPrintSemicolon
    \SetKwInOut{Input}{Input}\SetKwInOut{Output}{Output}
    \Input{A graph \(G = (V, E)\), a number of partitions \(k\), an imbalance \(\epsilon\), an optimizer \(opt\), a cut budget \(b\), and a number of shots \(s\)}
    \Output{A bitstring encoding an independent set of \(G\)}
    \(parts, S \gets \text{edgepartition}(G, k, \epsilon)\)\;
    \(S^\prime \gets \{\}\)\;
    \(\mathcal{N} \gets \{\}\)\;
    sort \(s_1, \ldots, s_{\lvert S\rvert} \in S\) by \(\lvert N(s_i)\rvert\)\;
    \For{\(s_i \in S\)}{
        \(\mathcal{N}^\prime \gets \mathcal{N} \cup (N(v_i) \setminus S)\)\;
        \If{\(\lvert \mathcal{N}^\prime\rvert \leq b\)}{
            \(N \gets N^\prime\)\;
            \(S^\prime \gets S^\prime \cup \{v_i\}\)
        }
    }
    \(\psi, \theta \gets \text{circuitgen}(G, parts, S^\prime)\)\;
    \(\theta^* \gets opt(\mathcal{L}, \psi, \theta)\)\;
    \(X \gets \text{sample}(\psi, \theta^*, s)\)\;
    \textbf{return} \(\argmax_{x \in X} \sum_{v\in V} x_v\)
\end{algorithm}

If the size of the separator is significant, this sort of sparsification likely isn't tractable. In this case, it likely isn't feasible to find a solution for the entire graph with only a single circuit. As an alternative to the sparsification approach, one could try an iterative strategy, where small subproblems are solved and recombined for find an overall solution. In this case, the local subproblems in question are defined by the subset of partitions connected by a single vertex \(s\) in the separator,  \(\{ V_i \mid i \leq k, \lvert N(s) \cap V_i\rvert > 0\}\).

Together with \(s\), these form an induced subgraph which can be solved to find a local part of the solution for the overall graph. The rest of the current solution remains fixed. Any mixers that may produce a conflict with the fixed solution is deactivated.

By iterating over all vertices in the separator, we generate a set of subproblems which covers the entire graph. Over the course of multiple sweeps, we can find and refine a global solution for the input graph. This approach is highly scalable, and can be used to find solutions for graphs with several thousand vertices. This algorithm is given as Alg. \ref{alg:iterative}.

\begin{algorithm}
    \caption{Deferred Constraint Quantum Divide and Conquer Algorithm (Iterative)}\label{alg:iterative}
    \DontPrintSemicolon
    \SetKwInOut{Input}{Input}\SetKwInOut{Output}{Output}
    \Input{A graph \(G = (V, E)\), a number of partitions \(k\), an imbalance \(\epsilon\), an optimizer \(opt\), a cut budget \(b\), a number of shots \(s\), and a number of iteratations \(iter\)}
    \Output{A bitstring encoding an independent set of \(G\)}
    \(x \gets [0] * \lvert V \rvert\)\;
    \(parts, S \gets \text{edgepartition}(G, k, \epsilon)\)\;
    sort \(s_1, \ldots, s_{\lvert S\rvert} \in S\) by \(\lvert N(s_i)\rvert\)\;
    \For{\(1 \leq i \leq iter\)}{
        \For{\(s_i \in S\)}{
            \If{\(\lvert\mathcal{N} \rvert \leq b\)}{
                \(local \gets \{p \in parts \mid \exists u \in p : s_i \in N(u)\}\)\;
                \(L \gets \{s_i\} \cup \{u \mid \forall p \in local, \forall u \in p\}\)\;
                \(L \gets \{v \in L \mid \exists u \in N(v) : x[u] = 1\} \)\;
                \(local \gets \{p \setminus L \mid p \in local\}\)\;
                \(\psi, \theta \gets \text{circuitgen}(G, local, \{s_i\}\setminus L)\)\;
                \(\theta^* \gets opt(\mathcal{L}, \psi, \theta)\)\;
                \(X \gets \text{sample}(\psi, \theta^*, s)\)\;
                \(x_{local} \gets \argmax_{x \in X} \sum_{v\in V} x_v\)
            }
        }
    }
    \textbf{return} \(x\)
\end{algorithm}

\section{Simulation Results}

In this section, we evaluate the performance of the DC-QDCA using the Pennylane library, evaluated on University of Delaware Caviness cluster. 
Our testing uses a collection of real world graphs from the Suite Sparse Matrix Collection \cite{davis2011university}, as well as a Random 3-Regular graph and a Connected Watts-Strogatz graph. The code for this project, the graphs we tested on, and a summary of the collected data will be available at \cite{githublink}. Our primary source of comparison will be the original QDCA implementation, found at \cite{dqcacode}.
Where optimal results were required in order to give an approximation ratio, we used the classical KaMIS library \cite{kamis}, while for the calculation of vertex separators we used an edge partition provided by the KaHIP library \cite{edgepartitioning2019}.

\subsection{Single Circuit Inputs}

For these tests, we compare directly with QDCA on a collection of small inputs for which we can construct a circuit covering the majority of the graph. For DC-DQVA, each of these graphs are partitioned into 2-4 parts. We are primarily concerned with number of inactive mixers and how that relates to the quality of the result. A large number of inactive mixers means a large portion of the graph is not being considered in the optimization, and further applications of the circuit may be required to find a solution. Both approaches in this test were given a cut budget of 6, allowing us to compare the difference in coverage given the same amount of overhead. Both random graphs where generated with 16 vertices, while the real world graphs are up to size 25.

In Fig. \ref{fig:small-comparison}, we see that DC-QDCA tends to produce circuits with a fraction of the inactive mixers in QDCA. In some cases, DC-QDCA is able to capture the entire graph with no inactive mixers whatsoever. Notably, there is not a perfect correlation between inactive mixers and final quality. 
\begin{figure}[t] 
    \centering
    \begin{subfigure}{0.4\textwidth}
        \centering
        \includegraphics[width=\textwidth]{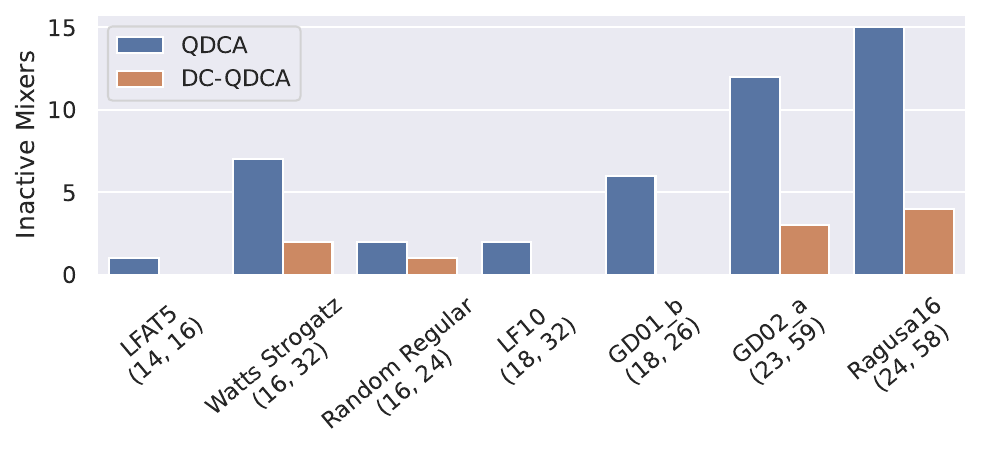}
        \caption{}
        \label{fig:mixers-small}
    \end{subfigure}\\
    \begin{subfigure}{0.4\textwidth}
        \centering
        \includegraphics[width=\textwidth]{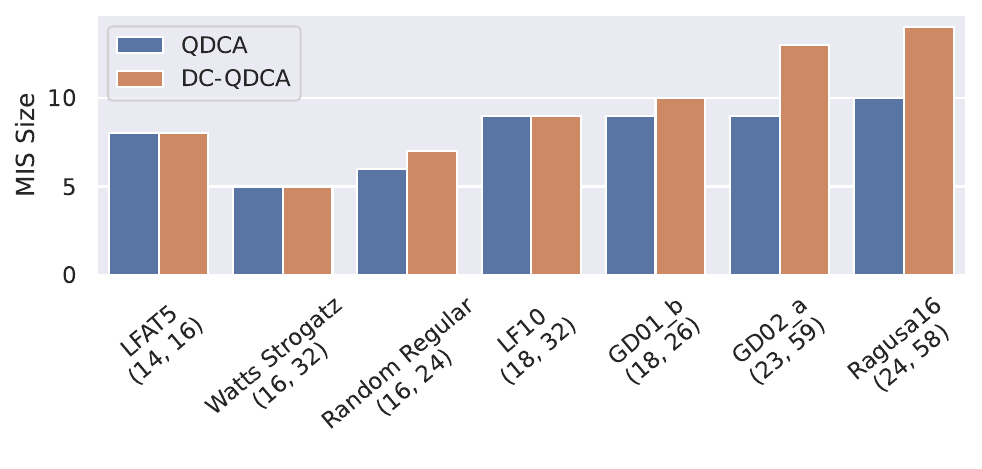}
        \caption{}
        \label{fig:small-res}
    \end{subfigure}
    \caption{Comparison of DC-QDCA with QDCA with respect to mixer activations on a number of small graphs with size \((\lvert V\rvert, \lvert E \rvert)\). Fig \ref{fig:mixers-small} shows the significant decrease in inactive mixers achieved by DC-QDCA while Fig \ref{fig:small-res} shows the accompanying increase in quality.}
    \label{fig:small-comparison}
\end{figure}

However, we can see that where the difference in inactive mixers is greatest, so too is the difference in quality. In fact, on each of these graphs, DC-QDCA was able to achieve an optimal result, whereas QDCA fell short on several, especially those where it had a great number of inactive mixers, such as Ragusa16 or GD02\_a.

\subsection{Iterative Approach}

To demonstrate the application of the iterative version of our algorithm, we look at a collection of 4 large real world graphs from the Suite Spares Matrix Collection. In particular, we look at both citation and reference networks, as a well as a network representation of the US power grid. The largest of these graphs, USpowergrid, requires a full 250 partitions in order to get the size of the largest partition small enough to be reasonably tractable.  A full list of parameters used for each individual graph is available at \cite{githublink}.

Our initial results in Fig. \ref{fig:iterative}, are highly promising, achieving an approximation ratio as high as \(97\%\). Moreover, an increase in size does not necessarily correlate with a decrease in the quality of this iterative method, as some of our best results were achieved for the largest graph USpowergrid.
\begin{figure}[t]
    \centering
    \includegraphics[width=0.35\textwidth]{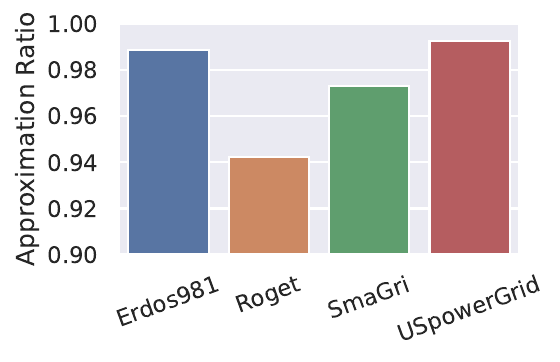}
    \caption{Approximation ratios for a collection of real world graphs using iterative DC-QDCA.}
    \label{fig:iterative}
\end{figure}

To more closesly examine the behavior of our iterative method, we demonstrate the quality of the solution overtime in Fig. \ref{fig:smagri-progress}. Here, we see the increase in quality over several hundred circuit applications over two full sweeps of the iterative method. Here we see a significant amount of improvement initially, which slowly tapers of as we approach the optimum, noted by the dotted line.
\begin{figure}[t]
    \centering
    \includegraphics[width=0.35\textwidth]{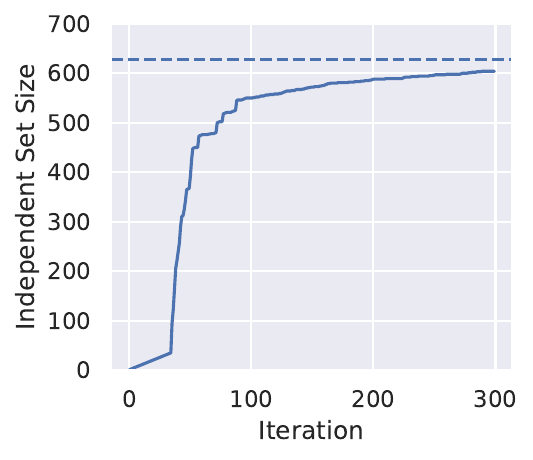}
    \caption{Shows the progress of iterative DC-QDCA on the SmaGri network over two sweeps of the iterative method.}
    \label{fig:smagri-progress}
\end{figure}
This behavior is highly promising, demonstrating that given access to fast quantum hardware, we could feasibly find high quality solutions for graphs with several thousand vertices, without actually requiring a device or system of devices which support several thousand qubits.

\section{Conclusion and Future Work} 

Scalability in quantum computing will remain one of the preeminent obstacles to achieving wide scale use in practice for the foreseeable future. In particular, it is a considerable barrier to demonstrating practical applications of quantum optimization. Previous attempts to address this problem include QDCA, which utilized a divide and conquer approach to the Maximimum Independent Set problem. We demonstrate an approach to achieve scalability in a distributed setting, utilizing edge partitioning and separator sparsification to lower communication costs and improve results over QDCA. Moreover, we introduced an iterative variant of our algorithm which allows us to find high quality solutions on real world graphs with several thousand vertices. Ongoing efforts are being made to show the broad applicability of these methods to other optimization problems, as well as expanding on existing results through improved heuristics.

\printbibliography 

\end{document}